\documentclass[12pt]{article}

\usepackage{amsmath,amssymb}
\usepackage{graphicx}

\makeatletter

\renewcommand\section{\@startsection{section}{1}{\z@}
                                   {-3.5ex \@plus -1ex \@minus -.2ex}
                                   {2.3ex \@plus .2ex}
                                   {\normalfont\large\bfseries}}
\renewcommand\subsection{\@startsection{subsection}{2}{\z@}
                                   {-3.25ex\@plus -1ex \@minus -.2ex}
                                   {1.5ex \@plus .2ex}
                                   {\normalfont\normalsize\bfseries}}
\renewcommand\subsubsection{\@startsection{subsubsection}{3}{\z@}
                                   {-3.25ex\@plus -1ex \@minus -.2ex}
                                   {1.5ex \@plus .2ex}
                                   {\normalfont\normalsize\bfseries}}
\renewcommand\paragraph{\@startsection{paragraph}{4}{\z@}
                                   {3.25ex \@plus1ex \@minus.2ex}
                                   {-1em}
                                   {\normalfont\normalsize\bfseries}}

\makeatother

\newcommand{\beq}{\begin{equation}}
\newcommand{\eeq}{\end{equation}}
\newcommand{\bea}{\begin{eqnarray}}
\newcommand{\eea}{\end{eqnarray}}

\newcommand{\SL}{{\rm SL}}
\newcommand{\SU}{{\rm SU}}

\newcommand{\Sp}{{\rm Sp}}
\newcommand{\Spin}{\rm Spin}

\newcommand{\so}{{\rm so}}
\newcommand{\symp}{{\rm sp}}
\newcommand{\R}{\mathbb R}
\newcommand{\Z}{\mathbb Z}

\newcommand{\id}{\hbox{1\kern-.27em l}}

\newcommand{\Tr}{{\rm Tr}}

\newcommand{\cC}{{\cal C}}
\newcommand{\cD}{{\cal D}}

\newcommand{\cG}{{\cal G}}
\newcommand{\cH}{{\cal H}}

\newcommand{\cL}{{\cal L}}

\newcommand{\cT}{{\cal T}}

\begin{document}

\pagestyle{empty}

\begin{center}

\vspace*{30mm}
{\Large  Automorphic properties of $(2, 0)$ theory on $T^6$}

\vspace*{30mm}
{\large M{\aa}ns Henningson}

\vspace*{5mm}
Department of Fundamental Physics\\
Chalmers University of Technology\\
S-412 96 G\"oteborg, Sweden\\[3mm]
{\tt mans@chalmers.se}     
     
\vspace*{30mm}{\bf Abstract:} 
\end{center}
We consider $ADE$-type $(2, 0)$ theory on a family of flat six-tori endowed with flat $\Sp (4)$ connections coupled to the $R$-symmetry. Our main objects of interest are the components of the `partition vector' of the theory. These constitute an element of a certain finite dimensional vector space, carrying an irreducible representation of a discrete Heisenberg group related to the 't~Hooft fluxes of the theory. Covariance under the $\SL_6 (\Z)$ mapping class group of a six-torus amounts to a certain automorphic transformation law for the partition vector, which we derive. Because of the absence of a Lagrangian formulation of $(2, 0)$ theory, this transformation property is not manifest, and gives useful non-trivial constraints on the partition vector. As an application, we derive a shifted quantization law for the spatial momentum of $(2, 0)$ theory on a space-time of the form $\R \times T^5$. This quantization law is in agreement with an earlier result based on the relationship between $(2, 0)$ theory and maximally supersymmetric Yang-Mills theory together with certain geometric facts about gauge bundles.
 
\newpage \pagestyle{plain}
 
 \section{Introduction}
The partition function is one of the most important quantities characterizing a quantum theory. In a Lagrangian formulation, the theory is defined by an action $S$, which is a functional of some fields $\phi$ over a (Euclidean) manifold $M_d$. The partition function can then be computed as a path integral over the space of all such classical (off shell) field configurations:
 \beq
 Z = \int \cD \phi \exp (-S) .
 \eeq
Possible refinements involve describing the domain of the functional integration more precisely (e.g. prescribing the topological class or the singularity structure of the fields) and/or insertions of field-dependent factors (observables) in the path-integral. The advantage of this formalism is that it keeps all space-time symmetries manifest. In an equivalent Hamiltonian formulation, space-time is factorized as $\R \times M_{d - 1}$, where the first factor is some chosen time-direction, and the second factor is a spatial manifold. The theory is then described in terms of a Hilbert space $\cH$, and the partition function is a generating function of a formal variable $t$ for the number of eigenstates with different eigenvalues for the Hamiltonian operator $H$: 
\beq
 Z = \Tr_{\cH} \exp (- t H) .
\eeq
 Also here it is possible to consider refinements, either involving different boundary conditions in the remaining spatial directions (which lead to different Hilbert spaces), or by classifying the states by their values of further commuting observable operators in addition to $H$. In this formalism, invariance under space-time symmetries that do not respect the choice of time direction is not manifest, and must be verified separately. The Hamiltonian expression for the partition function on $M_{d - 1}$ with formal variable $t$ agrees with the Lagrangian expression on $M_d = S^1 \times M_{d - 1}$ for a circle $S^1$ of circumference $t$.
  
In recent years, quantum theories that have no classical limit, and in particular no Lagrangian formulation, have become increasingly important. The best known class of such theories are those with $(2,0)$ superconformal symmetry in six space-time dimensions \cite{Witten95}. They obey an $ADE$-classification, just like the simply laced Lie algebras, but have no other discrete or continuous parameters. (However, in e.g. six-dimensional Minkowski space, they do have a moduli space of inequivalent vacua.) Their existence was first inferred by considerations in string theory and M-theory on higher-dimensional space-times including defects such as singularities or extended objects (branes). But it was also noted that upon compactification, the $(2, 0)$ theories may give rise to theories in  lower dimensions which do admit a Lagrangian formulation, notably maximally supersymmetric Yang-Mills theories. This perspective on the latter theories sheds considerable light on some of their more mysterious properties, such as strong-weak coupling duality ($S$-duality), which is not manifest in the Lagrangian formulation. By {\it reductio ad absurdum}, it also follows from the relationship to Yang-Mills theory that the $(2, 0)$ theories indeed do not admit a Lagrangian formulation \cite{Witten07}: Compactification of a $(2, 0)$ theory on a circle $S^1$ of radius $R$ gives rise to a five-dimensional maximally supersymmetric Yang-Mills theory with coupling constant $g = R^{1/2}$ and an action of the form $S = \frac{1}{g^2} \int_{M_5} d^5 x \cL_{\rm YM}$. The important point is that $S$ is inversely proportional to $R$. But a six-dimensional local Lagrangian $\cL_{(2, 0)}$ of $(2,0)$ theory would lead to a five-dimensional action $S = 2 \pi R \int d^5 x \cL_{(2, 0)}$  which is directly proportional to $R$. So there can be no such Lagrangian $\cL_{(2, 0)}$ for $(2, 0)$ theory.
  
 In this paper, we will consider $(2, 0)$ theory on a family of flat six-tori $T^6$. Our main object of interest is the dependence of the partition function of the theory on the geometry of $T^6$ and data related to the $\Sp (4)$ $R$-symmetry that is part of $(2, 0)$ superconformal symmetry. The set-up is described in detail in the next section. However, the theory does not have a single partition function, but rather a `partition vector' taking its values in a certain finite-dimensional complex vector space $V$ \cite{Witten98}\cite{Witten09}. A choice of time direction (which anyway is needed in the Hamiltonian formulation that we will use) gives a natural basis for this vector space. But in this formulation, six-dimensional covariance is certainly not manifest. It amounts to a certain automorphic transformation law of the partition vector under the $\SL_6 (\Z)$ mapping class group of $T^6$. This constitutes the main result of this paper.

Ultimately, one would of course like to determine the partition vector, at least implicitly, but this is well beyond the scope of the present paper. However, I would like to think that this goal is not completely unrealistic, and I hope to be able to continue this line of thought in the near future. The automorphic properties that are the focus of this paper should then be most important. Supersymmetry together with some analyticity properties imposes further strong constraints, and the Yang-Mills approximation gives useful boundary restrictions in the limit when one of the cycles of $T^6$ becomes small. 

In section three of the present paper, we will content ourselves with one small, but rather illuminating application, though: As discussed above, the partition function on $T^6 = S^1 \times T^5$ can also be viewed as pertaining to a space-time of the form $\R \times T^5$, where the first factor denotes time and the second factor is a spatial manifold. Applying the automorphic transformation law to the special case of transformations in which the time cycle $S^1$ is shifted by an arbitrary spatial cycle on $T^5$, we will recover a shifted quantization law for the spatial momentum $p$ on $T^5$. The result agrees with that of earlier considerations based on the relationship between this $(2, 0)$ theory in the case when $T^5 = T^4 \times S^1$ for a small $S^1$  and weakly coupled maximally supersymmetric Yang-Mills theory on a space-time of the form $\R \times T^4$ \cite{Henningson09}. Certain geometric facts about gauge bundles played an important role for the latter argument, and we find it gratifying that it can now be confirmed by the rather different, and more abstract, methods of the present paper.

\section{$(2, 0)$ theory on $T^6$}
The $(2, 0)$ theories obey an $ADE$-classification, i.e. they are in one-to-one correspondence with e.g. the simply laced Lie algebras. For a given $ADE$-type $\Phi$, we let the group $\cC$ of order $| \cC |$ denote the center of the corresponding simply connected compact Lie group $\cG$. The different possibilities are as follows:
\beq
\begin{array}{llll}
{\rm \Phi} & \cG & \cC & | \cC | \cr
\hline
A_{N-1} & \SU (N) & \Z / N \Z & N \cr
D_{2 k + 1} & \Spin (4 k + 2) & \Z / 4 \Z & 4 \cr
D_{4 k} & \Spin (8 k) & (\Z / 2 \Z) \times (\Z / 2 \Z) & 4 \cr
D_{4 k + 2} & \Spin (8 k + 4) & (\Z / 2 \Z) \times (\Z / 2 \Z) & 4\cr
E_6 & E_6 & \Z / 3 \Z & 3\cr
E_7 & E_7 & \Z / 2 \Z & 2\cr
E_8 & E_8 & 1 & 1 .
\end{array}
\eeq
In terms of the root lattice $\Gamma$ of $\cG$ and its dual $\Gamma^*$ (the weight lattice), the finite abelian group $\cC$ can be identified with
\beq
\cC \simeq \Gamma^* / \Gamma .
\eeq
The inner product on the root space $\Gamma \otimes \R$ thus endows $\cC$ with a non-degenerate symmetric bilinear $\R / \Z$-valued pairing 
\beq
c \cdot c^\prime \in \R / \Z
\eeq
for $c, c^\prime \in \cC$. For the cyclic cases $\cC \simeq \mathbb Z / N \mathbb Z$ we have
\beq
c \cdot c^\prime = \frac{1}{N} c c^\prime \in \frac{1}{N} \Z / \Z 
\eeq
(computed by first lifting $c$ and $c^\prime$ to $\Z$, multiplying, dividing by $N$ and finally reducing modulo $\Z$). For the remaining non-cyclic cases $\cC \simeq (\Z / 2 \Z) \times (\Z / 2 \Z)$ we have to distinguish between two cases: For $D_{4 k}$ we have the `triality' invariant product
\beq
\begin{array}{ccccc}
\cdot & (0, 0) & (1, 0) & (1, 1) & (0, 1) \cr
\hline
(0, 0) & 0 & 0 & 0 & 0 \cr
(1, 0) & 0 & 0 & {\scriptstyle \frac{1}{2}} & {\scriptstyle \frac{1}{2}} \cr
(1, 1) & 0 & {\scriptstyle \frac{1}{2}} & 0 & {\scriptstyle \frac{1}{2}} \cr
(0, 1) & 0 & {\scriptstyle \frac{1}{2}} & {\scriptstyle \frac{1}{2}} & 0 
\end{array} ,
\eeq
whereas for $D_{4 k + 2}$
\beq
\begin{array}{ccccc}
\cdot & (0, 0) & (1, 0) & (1, 1) & (0, 1) \cr
\hline
(0, 0) & 0 & 0 & 0 & 0 \cr
(1, 0) & 0 & {\scriptstyle \frac{1}{2}} & {\scriptstyle \frac{1}{2}} & 0 \cr
(1, 1) & 0 & {\scriptstyle \frac{1}{2}} & 0 & {\scriptstyle \frac{1}{2}} \cr
(0, 1) & 0 & 0 & {\scriptstyle \frac{1}{2}} & {\scriptstyle \frac{1}{2}} 
\end{array} .
\eeq

In six-dimensional Minkowski space $\R^{1, 5}$, a $(2, 0)$ theory is invariant under the $(2, 0)$ superconformal algebra ${\rm osp} (6, 2Ê|Ê4)$ \cite{Nahm}. The even subalgebra is $\so (6, 2) \oplus \symp (4)$, where the two terms denote the conformal algebra in six dimensions and the $R$-symmetry algebra respectively. The odd generators transform in the $(8, 4)$ representation of the even subalgebra, where $8$ is interpreted as a spinor representation of $\so (6, 2)$. (Because of $\so (6, 2)$ triality, the designation of its three inequivalent eight-dimensional representations as vector, spinor, and cospinor respectively is purely conventional.) Under the $\so (5, 1) \oplus \so (1, 1) \subset \so (6, 2)$ subalgebra of Lorentz and scale transformations, this representation decomposes as $8 = (4_s, {\scriptstyle \frac{1}{2}}) \oplus (4_c, - {\scriptstyle \frac{1}{2}})$, where the two terms denote a chiral and anti-chiral spinor of positive and negative scaling dimension respectively.
  
But instead of Minkowski space, we will consider a (Euclidean) six-torus
 \beq
 T^6 = \R^6 / \Lambda_6 ,
 \eeq
 where $\R^6$ is endowed with the standard inner product, and $\Lambda_6 \subset \R^6$ is a rank six lattice, which thus induces a flat metric $G$ on $T^6$. It is quite natural to couple the $R$-symmetry of a $(2, 0)$ theory to a flat $\Sp (4)$ connection over $T^6$. Such a connection is determined by its holonomies, which can be conjugated to a maximal torus subgroup $\cT$ of $\Sp (4)$. In terms of the root space $\Gamma_{\rm root} \otimes \R$ and the coroot lattice $\Gamma_{\rm coroot}$ of $\Sp (4)$, we have
 \beq \label{Cartan}
 \cT \simeq \Gamma_{\rm root} \otimes \R / \Gamma_{\rm coroot} .
\eeq
The holonomies can now be identified\footnote{Actually, we should also identify holonomies related by the action of the Weyl group $W \simeq \Z_2 \ltimes \Z_2^2$ of $\Sp (4)$.} with an element
 \beq
 \Theta \in H^1 (T^6, \cT) \simeq {\rm Hom} (\pi_1 (T^6), \cT) .
 \eeq
  
This setup partially breaks the $(2, 0)$ superconformal symmetry, but preserves the supertranslations subalgebra generated by six-dimensional translations, ordinary supersymmetries of positive scaling dimension, and the $R$-symmetry algebra. Our aim in the rest of the paper is to investigate the dependence of the theory on these geometric data, i.e. the metric $G$ and the holonomies $\Theta$.
 
 \subsection{The partition vector}
 As mentioned in the introduction, the partition function of the $(2, 0)$ theory is best thought of as an element of of a certain finite-dimensional complex vector space $V$ \cite{Witten98}\cite{Witten09}. This space is the representation space of the (up to unitary equivalence) unique unitary representation of a certain discrete Heisenberg group, as we will now explain. In addition to its central elements (which are represented as roots of unity), the Heisenberg group has elements $\Phi_u$ labeled by $u \in H^3 (T^6, \cC)$. These obey the commutation relations
 \beq
 \Phi_u \Phi_v = \Phi_v \Phi_u \exp \left( 2 \pi i \int_{T^6} u \cdot v \right) 
 \eeq
for $u, v \in H^3 (T^6, \cC)$. Here the symplectic (i.e. anti-symmetric and non-degenerate) product $u \cdot v \in H^6 (T^6, \R / \Z)$ is obtained by composing the symmetric inner product on $\cC$ with the anti-symmetric cup product on the rank-three cohomology. But to define the Heisenberg group, these commutation relations must be refined to a multiplication law 
\beq
 \Phi_u \Phi_v =  \pm \sqrt{\exp \left( 2 \pi i \int_{T^6} u \cdot v \right)} \Phi_{u + v} .
\eeq
The choice of signs for the square roots for different $u, v \in H^3 (T^6, \cC)$ are constrained by associativity, i.e. 
\beq
\Phi_u (\Phi_v \Phi_w) = (\Phi_u \Phi_v) \Phi_w 
\eeq
for $u, v, w \in H^3 (T^6, \cC)$. In general, the different choices are permuted by the ${\rm Aut} (\Lambda_6) \simeq SL_6 (\Z)$ mapping class group of $T^6$, which acts on $H^3 (T^6, \cC)$ by permutations\footnote{When $\cC \simeq \Z / N \Z$ for $N$ odd, i.e. the $(2, 0)$ theory is of type $A_{N -1}$ for $N$ odd, $E_6$, or $E_8$, there is actually an ${\rm Aut} (\Lambda_6)$-invariant choice of multiplication law, namely  $\Phi_u \Phi_v = \exp \left( 2 \pi i \frac{1-N}{2} \int_{T^6} u \cdot v \right) \Phi_{u + v}$, which obeys both the commutation relations and the associativity constraint.}.

To construct an irreducible representation of this Heisenberg group in a vector space $V$, we choose a decomposition (polarization) of $H^3 (T^6, \cC)$
\bea \label{polarization}
H^3 (T^6, \cC) & = & F \oplus G \cr
u & = & f + g 
\eea
such that
\beq
f \cdot f^\prime = g \cdot g^\prime = 0
\eeq
for all $f, f^\prime \in F$ and $g, g^\prime \in G$ \footnote{Hopefully the use of the symbol $G$ for both the metric on $T^6$ and this subspace of $H^3 (T^6, \cC)$ will not cause any confusion.}. The order of the groups $F$ and $G$ are then
\bea
| F | & = & | G | \cr
& = & \sqrt{| H^3 (T^6, \cC) |} \cr
& = & | \cC |^{10} .
\eea
In the vector space $V$ there is now a unique ray, represented by a non-zero vector $\Psi_0 \in V$ invariant under the (commuting) elements $\Phi_g$ for all $g \in G \subset H^3 (T^6, \cC)$, i.e.
\beq
\Phi_g \Psi_0 = \Psi_0 .
\eeq
An orthonormal basis of $V$ is given by $\{ \Psi_f \}_{f \in F}$, where
\beq
\Psi_f = \Phi_f \Psi_0 ,
\eeq
so $\dim_{\mathbb C} V = | \cC |^{10}$. We refer to $f \in F$ as the (discrete abelian) 't~Hooft flux of the $(2, 0)$ theory.  It labels the components $Z_f (G | \Theta)$ of the partition vector $Z (G | \Theta)$ with respect to the basis $\{\Psi_f \}_{f \in F}$ of $V$:
\beq
Z (G | \Theta) = \sum_{f \in F} Z_f (G | \Theta) \Psi_f .
\eeq

A concrete way of choosing a polarization of $H^3 (T^6, \cC) = H^3 (T^5, \Z) \otimes \cC$ is induced by a (not necessarily orthogonal) decomposition of the lattice $\Lambda_6$:
\beq \label{Lambda-decomposition}
\Lambda_6 = \lambda_0 \otimes \Z \oplus \Lambda_5 ,
\eeq
where $\lambda_0$ is a primitive element and $\Lambda_5$ a rank five sub-lattice of $\Lambda_6$. This induces an orthogonal decomposition
\beq
\R^6 = \R_{\rm Time} \oplus \R^5_{\rm Space} ,
\eeq
where $\R^5_{\rm Space} = \Lambda_5 \otimes \R$ and $\R_{\rm Time} = (\R^5_{\rm Space})^\perp$. The six-torus can then be factorized as
\beq
T^6 = S^1 \times T^5 ,
\eeq
where $S^1 = \lambda_0 \otimes (\R / \Z)$ and $T^5 = \Lambda_5 \otimes (\R / \Z)$. The polarization (\ref{polarization}) with
\bea
F & \simeq & H^0 (S^1, \cC) \otimes H^3 (T^5, \cC) \cr
& \simeq & H^3 (T^5, \cC)
\eea
and
\bea
G & \simeq & H^1 (S^1, \cC) \otimes H^2 (T^5, \cC) \cr
& \simeq & H^2 (T^5, \cC)
\eea
now amounts to the K\"unneth isomorphism.

\subsection{Automorphic properties}
Relative to a choice of basis of $\Lambda_6 \simeq H^1 (T^6, \Z)$, the metric $G$ on $T^6$ and the $\Sp (4)$ holonomies $\Theta$ can be represented as a real, symmetric and positive definite $6 \times 6$ matrix, and a six-dimensional vector with values in the maximal torus $\cT$ respectively. But the action of the ${\rm Aut} (\Lambda_6) \simeq \SL_6 (\Z)$ mapping class group of $T^6$ imposes discrete identifications on these data. 

To extract the implications of these identifications for the components $Z_f (G | \Theta)$ of the partition vector, we need to understand the action of ${\rm Aut} (\Lambda_6)$ on the vector space $V$. The action (by permutation) of $\sigma \in {\rm Aut} (\Lambda_6)$ on $H^3 (T^6, \cC)$ defines an element of the symplectic group $\Sp (H^3 (T^6, \cC))$ in the sense that it leaves the symplectic product invariant:
\beq
\sigma u \cdot \sigma v = u \cdot v 
\eeq
for all $u, v \in H^3 (T^6, \cC)$. In terms of the polarization (\ref{polarization}), this action can be expressed as a matrix of maps 
\beq \label{sigma-hat}
\hat{\sigma} = \left(
\begin{matrix}
A & B \cr
C & D 
\end{matrix}
\right) : \left(
\begin{matrix}
F \rightarrow F & G \rightarrow F \cr
F \rightarrow G & G \rightarrow G 
\end{matrix}
\right)
\eeq
obeying
\bea
A f \cdot C f^\prime + C f \cdot A f^\prime & = & 0 \cr
B g \cdot D g^\prime + D g \cdot B g^\prime & = & 0 \cr
A f \cdot D g^\prime + C f \cdot B g^\prime & = & f \cdot g^\prime
\eea
for all $f, f^\prime \in F$ and $g, g^\prime \in G$. 
 
We begin by determining the action of $\sigma \in {\rm Aut} (\Lambda_6)$ on $\Psi_0 \in V$. The vector $\sigma \Psi_0 \in V$ should obey
\bea
\Phi_{B g + D g} \, \sigma \Psi_0 & = & \sigma \Phi_g \sigma^{-1} \sigma \Psi_0 \cr
& = & \sigma \Phi_g \Psi_0 \cr
& = & \sigma \Psi_0 ,
\eea
i.e. it should be invariant under $\Phi_{B g + D g}$, for all $g \in G$. The solution is
\bea
\sigma \Psi_0 & = & c_\sigma \sum_{g \in G} \Phi_{B g + D g} \Psi_0 \cr
& = & c_\sigma \sum_{g \in G} \sqrt{\exp \left(- 2 \pi i \int_{T^6} B g \cdot D g \right)} \Phi_{B g} \Phi_{D g} \Psi_0 \cr
& = & c_\sigma \sum_{g \in G} \sqrt{\exp \left(- 2 \pi i \int_{T^6} B g \cdot D g \right)} \Psi_{B g} , 
\eea
where the constant $c_\sigma$ is determined up to a complex phase by the requirement that $\sigma \Psi_0$ be normalized. We see that the modulus $| c_\sigma |$ only depends on $B$ and $D$. We can now compute the action of $\sigma \in {\rm Aut} (\Lambda_6)$ on an arbitrary basis vector $\Psi_f = \Phi_f \Psi_0 \in V$: 
\bea
\sigma \Psi_f & = & \sigma \Phi_f \Psi_0 \cr
& = & \sigma \Phi_f \sigma^{-1} \sigma \Psi_0 \cr
& = & \Phi_{A f + C f} c_\sigma \sum_{g \in G} \Phi_{B g + D g} \Psi_0 \cr
& = &  c_\sigma \sum_{g \in G} \sqrt{\exp \left(2 \pi i \int_{T^6} 2 C f \cdot B g - A f \cdot C f - B g \cdot D g \right)} \Phi_{A f + B g} \Phi_{C f + D g} \Psi_0 \cr
& = &  c_\sigma \sum_{g \in G} \sqrt{\exp \left(2 \pi i \int_{T^6} 2 C f \cdot B g - A f \cdot C f - B g \cdot D g \right)} \Psi_{A f + B g} .
\eea

For a general choice of maps $A$, $B$, $C$, and $D$, subject only to the condition that they define an element $\hat{\sigma}$ of the symplectic group $\Sp (H^3 (T^6, \cC))$, there is no canonical choice of square roots in this formula. But for a transformation induced from an element $\sigma \in {\rm Aut} (\Lambda_6)$, not only $2 C f \cdot B g$  but also $A f \cdot C f$ and $B g \cdot D g$ are divisible by $2$ in $H^6 (T^6, \R / \Z)$ in a natural way. For $A f \cdot C f$, this can be seen by expanding $f$ as a linear combination of monomials of the form $e^1 \cup e^2 \cup e^3$, where $e^1, e^2, e^3 \in H^1 (T^6, \cC)$. Since $A f_1 \cdot C f_2 + A f_2 \cdot C f_1 = 2 A f_1 \cdot C f_2$, the cross-terms in $A f \cdot C f$ are divisible by $2$ in a natural way. And the diagonal terms of the form $A (e^1 \cup e^2 \cup e^3) \cdot C (e^1 \cup e^2 \cup e^3)$ vanish identically. The reasoning for $B g \cdot D g$ is completely analogous. We may thus, without ambiguity, write
\beq
\sigma \Psi_f = c_\sigma \sum_{g \in G} \exp \left(2 \pi i \int_{T^6} C f \cdot B g - {\scriptstyle \frac{1}{2}} A f \cdot C f - {\scriptstyle \frac{1}{2}} B g \cdot D g \right)  \Psi_{A f + B g} .
\eeq
The corresponding relationship between the partition vectors $Z_f (G | \Theta)$ and $Z_f ({}^\sigma G | {}^\sigma \Theta)$ for geometric data $G$, $\Theta$ and ${}^\sigma G$ and  ${}^\sigma \Theta$ related by $\sigma \in {\rm Aut} (\Lambda_6)$ is
\bea \label{automorphic-transformation}
Z_f (G | \Theta) & =  & c_\sigma \sum_{g \in G} \exp \left(2 \pi i \int_{T^6} \left(C f \cdot B g - {\scriptstyle \frac{1}{2}} A f \cdot C f - {\scriptstyle \frac{1}{2}} B g \cdot D g \right) \right) \cr
& & \times Z_{A f + B g} ({}^\sigma G | {}^\sigma \Theta) .
\eea
This automorphic transformation property is the main result of this paper.

\subsection{Examples}
When $\sigma$ belongs to the subgroup of $ {\rm Aut} (\Lambda_6)$ of elements for which $B = 0$, we have $c_\sigma = e^{i \phi} | G |^{-1}  = e^{i \phi} | \cC |^{-10}$ for some real phase $\phi$ so that
\beq
Z_f (G | \Theta) = e^{i \phi} \exp \left( 2 \pi i \int_{T^6} \left( - {\scriptstyle \frac{1}{2}} A f \cdot C f \right) \right) Z_{A f} ({}^\sigma G | {}^\sigma \Theta) .
\eeq
In particular, if also $C = 0$ so that $\sigma$ respects the decomposition $H^3 (T^6, \cC) = F \oplus G$, the transformation acts (up to a common phase) by permutation on $f$:
\beq
Z_f (G | \Theta) = e^{i \phi} Z_{A f} ({}^\sigma G | {}^\sigma \Theta) .
\eeq
Another special case, which will be important in the next section, is when $A = \id$ (and still $B = 0)$ so that the components transform with $f$-dependent phases:
\beq \label{special-automorphic}
Z_f (G | \Theta) = e^{i \phi} \exp \left( 2 \pi i \int_{T^6} \left( - {\scriptstyle \frac{1}{2}} f \cdot C f \right) \right) Z_f ({}^\sigma G | {}^\sigma \Theta) .
\eeq

For another example, suppose that $F$ and $G$ are further decomposed as
\bea
F & = & F_0 \oplus F_1 \cr
G & = & G_0 \oplus G_1, 
\eea
with 
\beq
F_0 \simeq G_0 ,
\eeq 
$| F_0 | = | G_0 | = | \cC |^6$, $| F_1 | = | G_1 | = | \cCÊ|^4$,
and 
\bea
f_0 \cdot g_1 & = & 0 \cr
f_1 \cdot g_0 & = & 0
\eea
for all $f_0 \in F_0$, $f_1 \in F_1$, $g_0 \in G_0$, and $g_1 \in G_1$. We consider the subgroup of $ {\rm Aut} (\Lambda_6)$ of elements such that
\bea
A (f_0 + f_1) & = & a f_0 + f_1 \cr
B (g_0 + g_1) & = & b g_0 \cr
C (f_0 + f_1) & = & c f_0 \cr
D (g_0 + g_1) & = & d g_0 + g_1
\eea
for some integers $a$, $b$, $c$, and $d$ subject to the requirement
\beq
\left( \begin{matrix}a & b \cr c & d \end{matrix} \right)  \in \SL_2 (\Z) .
\eeq
(In the right hand side, $b g_0$ and $c f_0$ should be interpreted as elements of $F_0$ and $G_0$ respectively.) We then have
\bea
Z_{f_0 + f_1} (G | \Theta) & = & c_\sigma | \cC |^4 \sum_{g_0 \in G_0} \exp \left(2 \pi i \int_{T^6} \left( c f_0 \cdot b g_0 - {\scriptstyle \frac{1}{2}} a f_0 \cdot c f_0 - {\scriptstyle \frac{1}{2}} b g_0 \cdot d g_0 \right) \right) \cr
& & \times Z_{a f_0 + b g_0 + f_1} ({}^\sigma G | {}^\sigma \Theta) .
\eea
Consider first the transformation $\sigma = S$, i.e.
\beq
\left( \begin{matrix}a & b \cr c & d \end{matrix} \right) = \left( \begin{matrix}0 & 1 \cr -1 & 0 \end{matrix} \right) .
\eeq
With $c_S = e^{i \phi} | \cC |^{-7}$ we have
\beq
Z_{f_0 + f_1} (G | \Theta) = e^{i \phi} | \cC |^{-3} \sum_{g_0 \in G_0} \exp \left(2 \pi i \int_{T^6} \left( - f_0 \cdot g_0 \right) \right) Z_{g_0 + f_1} ({}^S G | {}^S \Theta) .
\eeq
Repeating this transformation gives
\bea
Z_{f_0 + f_1} (G | \Theta) & = & e^{2 i \phi} | \cC |^{-6} \sum_{g_0 \in G_0} \sum_{g_0^\prime \in G_0} \exp \left(2 \pi i \int_{T^6} \left( - f_0 \cdot g_0 - g_0 \cdot g_0^\prime \right) \right)  \cr
& & \times Z_{g_0^\prime + f_1} ({}^{S^2} G | {}^{S^2} \Theta) \cr
& = & e^{2 i \phi} Z_{f_0 + f_1} ({}^{S^2} G | {}^{S^2} \Theta) .
\eea
So with $e^{i \phi}$ a square root of unity, we have the relation 
\beq
S^2 = \id .
\eeq 
Consider next the transformation $\sigma = S T$, i.e.
\beq
\left( \begin{matrix}a & b \cr c & d \end{matrix} \right) = \left( \begin{matrix}0 & 1 \cr -1 & -1 \end{matrix} \right) .
\eeq
With $c_{S T} = e^{i \phi} | \cC |^{-7}$ we have
\bea
Z_{f_0 + f_1} (G | \Theta) & = & e^{i \phi} | \cC |^{-3} \sum_{g_0 \in G_0} \exp \left(2 \pi i \int_{T^6} \left( - f_0 \cdot g_0 - {\scriptstyle \frac{1}{2}} g_0 \cdot g_0 \right) \right) \cr
& & \times Z_{g_0 + f_1} ({}^{S T} G | {}^{S T} \Theta) .
\eea
The two factors of $g_0$ in the second term in the exponent should be interpreted as elements of the isomorphic spaces $G_0$ and $F_0$ respectively, so their symplectic product is not identically zero. Instead we could think of the product as being symmetric in the two factors. Repeating this transformation twice gives
\bea
Z_{f_0 + f_1} (G | \Theta) & = & e^{3 i \phi} | \cC |^{-9} \sum_{g_0 \in G_0} \sum_{g_0^\prime \in G_0} \sum_{g_0^{\prime \prime} \in G_0} \exp \left(2 \pi i \int_{T^6} \left( - f_0 \cdot g_0 - {\scriptstyle \frac{1}{2}} g_0 \cdot g_0 \right. \right. \cr
& & \left. \left. - g_0 \cdot g_0^\prime - {\scriptstyle \frac{1}{2}} g_0^\prime \cdot g_0^\prime - g_0^\prime \cdot g_0^{\prime \prime} - {\scriptstyle \frac{1}{2}} g_0^{\prime \prime} \cdot g_0^{\prime \prime} \right) \right) \cr
& & \times Z_{g_0^{\prime \prime} + f_1} ({}^{(S T)^3} G | {}^{(S T)^3} \Theta) \cr
& = & e^{3 i \phi} | \cC |^{-9} \sum_{g_0 \in G_0} \sum_{g_0^\prime \in G_0} \sum_{g_0^{\prime \prime} \in G_0}  \exp \left(2 \pi i \int_{T^6} \left( {\scriptstyle \frac{1}{2}} f_0 \cdot f_0 - {\scriptstyle \frac{1}{2}} g_0^{\prime \prime} \cdot g_0^{\prime \prime} \right. \right. \cr
& & \left. \left. - {\scriptstyle \frac{1}{2}} (g_0 + f_0 + g_0^\prime) \cdot (g_0 + f_0 + g_0^\prime) + g_0^\prime \cdot (f_0 - g_0^{\prime \prime}) \right) \right) \cr
& & \times Z_{g_0^{\prime \prime} + f_1} ({}^{(S T)^3} G | {}^{(S T)^3} \Theta) \cr
& = & e^{3 i \phi} Z_{f_0 + f_1} ({}^{(S T)^3} G | {}^{(S T)^3} \Theta) ,
\eea
where in the last line we have used that
\beq
\sum_{g_0 \in G_0} \exp \left( 2 \pi i \int_{T^6} {\scriptstyle \frac{1}{2}} (g_0 + f_0 + g_0^\prime) \cdot (g_0 + f_0 + g_0^\prime) \right) = | \cC |^3 .
\eeq
(This can be derived by decomposing $G_0$ further as a sum of two subgroups of order $| \cC |^3$, such that all elements of either of the two terms have vanishing products with each other.) So with $e^{i \phi}$ a cubic root of unity, we have the relation 
\beq
(S T)^3 = \id . 
\eeq
We have thus recovered the well-known presentation of the group $\SL_2 (\Z)$ in terms of generators $S$ and $S T$ and relations.

\section{Application to the quantization of momentum}
As an application of the above results, we will consider the quantization law of the spatial momentum $p$ of $(2, 0)$ theory on a space-time of the form 
\beq
M^{1, 5} = \R \times T^5 , 
\eeq
where the two factors denote time and a flat spatial five-torus with metric $g$ (given by a real, symmetric positive definite $5 \times 5$ matrix) respectively. The $R$-symmetry is coupled to a flat $\Sp (4)$ connection over $T^5 = \Lambda_5 \otimes \R / \Z$, which we identify via its holonomies with an element $\theta \in H^1 (T^5, \cT)$. (Here $\cT$ is a maximal torus subgroup of $\Sp (4)$ as described in section two.) 

We consider the components of the partition vector in a Hamiltonian formalism:
\beq
Z_f (t, x, g | \theta_0, \theta) = {\rm Tr}_{\cH_{f, \theta}} \left( e^{- t H + i x P + i \theta_0 J} \right) ,
\eeq
where $H$, $P$, and $J$ are the Hamiltonian, spatial momentum operators and $\symp (4)$ $R$-symmetry generators respectively, and $t \in \R$, $x \in \Lambda_5 \otimes \R$, and $\theta_0 \in \cT$ are some formal parameters. The trace is taken over the Hilbert space $\cH_{f, \theta}$ of states with 't~Hooft flux $f \in F = H^3 (T^5, \cC)$ and spatial $R$-symmetry twist $\theta \in H^1 (T^5, \cT)$. But (after analytic continuation to Euclidean time) this partition vector also admits an interpretation in terms of $(2, 0)$ theory on a six-torus 
\beq
T^6 =  S^1 \times T^5
\eeq
 with a flat metric $G$, determined by $g$, $t$, and $x$, and a flat $\Sp (4)$ connection $\Theta$, determined by $\theta_0$ and $\theta$. In terms of the decomposition (\ref{Lambda-decomposition}), $x$ and $t$ are the orthogonal projections of $\lambda_0$ on  $\R^5_{\rm Space} = \Lambda_5 \otimes \R$ and $\R_{\rm Time} = (R^5_{\rm Space})^\perp$ respectively, and $\Theta$ is related to $\theta$ and $\theta_0$ by the K\"unneth isomorphism
\bea
H^1 (T^6, \cT) & \simeq & H^0 (S^1, \cT) \otimes H^1 (T^5, \cT) \oplus H^1 (S^1, \cT) \otimes H^0 (T^5, \cT) \cr
& \simeq & H^1 (T^5, \cT) \oplus H^0 (T^5, \cT) \cr
\Theta & = & \theta + \theta_0 .
\eea

The crucial point is now that the parameter $x \in \Lambda_5 \otimes \R \simeq H_1 (T^5, \R)$ is periodic: The transformation $G \mapsto {}^\sigma G$ defined by
\bea
x & \mapsto & x + \Delta x \cr
g & \mapsto & g \cr
t & \mapsto & t
\eea
for $\Delta x \in \Lambda_5 \simeq H_1 (T^5, \Z)$ leads to an isomorphic six-torus $T^6$. The induced action $\Theta \mapsto {}^\sigma \Theta$ is given by
\bea
\theta_0 & \mapsto & \theta_0 + \int_{\Delta x} \theta \cr
\theta & \mapsto & \theta .
\eea
And the induced action $u \mapsto {}^\sigma u$ can be described as in (\ref{sigma-hat}) with
\beq
\left(
\begin{matrix}
A & B \cr
C & D 
\end{matrix}
\right) = \left(
\begin{matrix}
\id & 0 \cr
- \iota_{\Delta x} & \id 
\end{matrix}
\right) ,
\eeq
where the contraction map
\beq
\iota_{\Delta x} : H^3 (T^5, \cC) = F \rightarrow G = H^2 (T^5, \cC)
\eeq
is defined by the equation
\beq
\int_{S} \iota_{\Delta x} f = \int_{\Delta x \times S} f
\eeq
for any $f \in F = H^3 (T^5, \cC)$ and an arbitrary two-cycle $S \in H_2 (T^5, \Z)$. It now follows from the special case (\ref{special-automorphic}) of the general formula (\ref{automorphic-transformation}) that
\beq
Z_f (G | \Theta) = \exp \left( 2 \pi i \int_{T^6} {\scriptstyle \frac{1}{2}} f \cdot \iota_{\Delta x} f \right) Z_f ({}^\sigma G | {}^\sigma \Theta) .
\eeq
Defining the class $\rho \in H^1 (T^5, \R / \Z)$, which depends only on $f \in H^3 (T^5, C)$, by the equation
\beq
\int_{\Delta x} \rho = \int_{T^5} {\scriptstyle \frac{1}{2}} f \cdot \iota_{\Delta x} f
\eeq
for all $\Delta x \in H_1 (T^5, \Z)$, this can be written as
\beq
Z_f (G | \Theta) = \exp \left( 2 \pi i \int_{\Delta x} \rho \right) Z_f ({}^\sigma G | {}^\sigma \Theta) .
\eeq

The interpretation of this result is clearer if we expand $Z_f (G | \Theta)$ in a Fourier series, replacing the continuous periodic variable $\theta_0 \in \cT$ with a discrete variable $w \in \Gamma_{\rm weight}$. Here $\Gamma_{\rm weight}$ is the weight lattice of $\Sp (4)$, i.e. the dual of the coroot lattice that appeared in the definition (\ref{Cartan}) of the maximal torus $\cT$. We thus write
\beq
Z_f (G | \theta, \theta_0) = \sum_{w \in \Gamma_{\rm weight}} Z_f^w (G | \theta) \exp \left(2 \pi i \theta_0 \cdot w \right) ,
\eeq
with some coefficients $ Z_f^w (G | \theta)$, which we interpret as the partition vector pertaining to the quantum states of $\Sp (4)$ weight $w$. In terms of these coefficients, the transformation law reads
\beq
Z_f^w (G | \theta) = \exp \left( 2 \pi i \int_{\Delta x} (\theta \cdot w + \rho) \right) Z_f^w ({}^\sigma G | \theta) .
\eeq
But from the Hamiltonian formulation, we expect that the partition function for quantum states of spatial momentum $p \in H^1 (T^5, \R)$ should obey
\beq
Z_f^w (G | \theta) = \exp \left( 2 \pi i \int_{\Delta x} p \right) Z_f^w ({}^\sigma G | \theta) .
\eeq
So we have deduced that the spectrum of $p$ obeys the shifted quantization law
\beq \label{p-shift}
p - \theta \cdot w - \rho  \in H^1 (T^5, \Z) \simeq \Lambda_5^* .
\eeq

The term $\theta \cdot w$ in the shift of the spatial momentum is what one would expect for quantum states of $\Sp (4)$ weight $w$ in a configuration with a spatial $\Sp (4)$ connection $\theta$. The term $\rho$, which as we have seen is bilinear in the 't~Hooft flux $f$, is maybe more surprising. But it has in fact be derived before by considerations based on the relationship between $(2, 0)$ theory on $\R \times T^5$ with $T^5 = T^4 \times S^1$ for a small $S^1$ and weakly coupled maximally supersymmetric Yang-Mills theory on $\R \times T^4$ \cite{Henningson09} : The Chern class (instanton number) $k$ of the gauge bundle over $T^4$ is then interpreted as the fifth component of the spatial momentum $p$ of the $(2, 0)$ theory. We also have the K\"unneth isomorphism
\bea
H^3 (T^5, \cC) & = & H^2 (T^4, \cC) \oplus H^3 (T^4, \cC) \cr 
f & = & m + e ,
\eea
where $m$ and $e$ are known as the magnetic and electric 't~Hooft fluxes of the Yang-Mills theory respectively. The magnetic 't~Hooft flux $m$ determines, together with the Chern class $k$, the topological class of the gauge bundle. The electric 't~Hooft flux $e$ determines the transformation properties of a state under gauge transformations that are `large' (i.e. not homotopic to the identity) in the sense that their parameters represent non-trivial classes of the fundamental group of the gauge group \cite{tHooft}. It is well-known (see e.g. \cite{Vafa-Witten} for an intuitive explanation) that $k$ is shifted away from integrality by an expression bilinear in $m$:
\beq
k - {\scriptstyle \frac{1}{2}} m \cdot m \in H^4 (T^4, \Z) .
\eeq
Somewhat less familiar is that in a situation with non-trivial $m$, a spatial translation by a cycle of $T^4$ is equivalent to a `large' gauge transformation \cite{Henningson09}. Indeed, a bundle with non-trivial magnetic 't~Hooft flux $m$ over a two-torus $T^2$ may be constructed by gluing together the ends of a cylinder $I \times S^1$ with a twist given by a gauge transformation which is large along the $S^1$ direction \cite{Witten00}. Translation along the $T^2$-cycle that originated from the interval $I$ thus amounts to a gauge transformation that is large along the $T^2$-cycle that originated from the circle $S^1$. A state of non-trivial electric 't~Hooft flux $e$ would thus transform with a non-trivial  phase factor, which can be interpreted as a shift away from the familiar integrality of the corresponding component of the four-dimensional spatial momentum $\tilde{p}$ by an expression linear in both $m$ and $e$:
\beq
\tilde{p} - m \cdot e \in H^1 (T^4, \Z) .
\eeq
In terms of the 't~Hooft flux $f$ and the five-dimensional spatial momentum $p$ of $(2, 0)$ theory, these results can be summarized as the second term in (\ref{p-shift}). But it is gratifying to be able to derive them directly from the postulated automorphic properties of $(2, 0)$ theory, without any reference to the low-energy effective Yang-Mills theory and the geometry of gauge bundles.

\vspace*{5mm}
This research was supported by the G\"oran Gustafsson foundation and the Swedish Research Council.

 \end{document}